# Experimental demonstration of an anisotropic exceptional point

Kun Ding[1], Guancong Ma[1,2†], Z. Q. Zhang[1], and C. T. Chan[1★]

[1]*Department of Physics and Institute for Advanced Study, Hong Kong University of Science and Technology, Clear Water Bay, Kowloon, Hong Kong*

[2]*Department of Physics, Hong Kong Baptist University, Kowloon Tong, Hong Kong*

† Email: phgcma@hkbu.edu.hk

★ Email: phchan@ust.hk

**Abstract**

Exceptional points (EPs) associated with a square-root singularity have been found in many non-Hermitian systems. In most of the studies, the EPs found are isotropic meaning that the same singular behavior is obtained independent of the direction from which they are approached in the parameter space. In this work, we demonstrate both theoretically and experimentally the existence of an anisotropic EP in an acoustic system that shows different singular behaviors when the anisotropic EP is approached from different directions in the parameter space. Such an anisotropic EP arises from the coalescence of two square-root EPs having the same chirality.

*Introduction.* In recent years, the existence of exceptional points (EPs) has been demonstrated in a variety of non-Hermitian systems [1-3]. Many useful applications of the physics associated with EPs have emerged, including unidirectional light propagation [4-6], sensors [7-9], coherent perfect absorption [10,11], single-mode lasers [12,13], loss-induced revival of lasing [14], unusual beam dynamics [15,16] and others [17-23]. An EP can be associated with a fractional topological charge [24-27] and many interesting phenomena related to EPs are rooted in their topological characteristics. For example, the most common and simplest form of EPs are associated with a square-root singularity and can be considered as carrying a topological charge of $\pm 1/2$, with the sign indicating the chirality of the EP [28-32]. More complex phenomena associated with EPs have been studied recently including a ring of EPs [33] and Fermi arcs [34,35]. In particular, it has recently been shown that the coalescence of multiple EPs can lead to higher-order EPs, which possess fractional winding numbers [31,36-38]. Higher-order EPs have been experimentally demonstrated in both acoustics [31] and photonics [8].

Most of the EPs studied so far are isotropic in the sense that the same singular behavior is obtained when the EP is approached from different directions in the parameter space. However, non-Hermitian systems carrying anisotropic EPs can be conceived and in those systems, different singular behaviors are obtained when the EP is approached from different directions in the parameter space. Such an EP can be realized when two square-root EPs having the same chirality coalesce and has been alternately referred to as an order-1 EP [31] or a hybrid EP [35]. It is worth mentioning that the coalescence of two EPs with opposite chiralities results in a diabolic point [39].

In this paper, we demonstrate both theoretically and experimentally the existence of a hybrid EP in an acoustic system with coupled cavities. We will use the coupling constant and the asymmetric loss of the two cavities as two tunable external parameters to identify the existence of a hybrid EP. The eigenvalues and phase rigidity [29] near the EP show different singular behaviors as the EP is approached by varying the two parameters.

*Conditions for achieving a hybrid EP.* Without loss of generality, let us consider a system composed of two cavities having the same resonant frequency $\omega_0$ and coupled through a coupling constant $\kappa$. The Hamiltonian of this system can be written as

$$H = \begin{pmatrix} \omega_0 - i\Gamma_0 & \kappa \\ \kappa & \omega_0 - i\Gamma \end{pmatrix}, \tag{1}$$

where $\Gamma_0$ denotes the intrinsic loss of the resonant state in Cavity A and $\Gamma = \Gamma_0 + \Delta\Gamma$ with $\Delta\Gamma$ representing the additional tunable loss of the resonant state in Cavity B. The eigenfrequencies $\tilde{\omega}_\pm$ and right eigenvectors $\tilde{\phi}_\pm^R$ of Eq. (1) take the following

forms:

$$\tilde{\omega}_\pm = \omega_0 - i\Gamma_{av} \pm \frac{1}{2}\sqrt{4\kappa^2 - (\Delta\Gamma)^2}, \tag{2}$$

$$\tilde{\phi}_\pm^R = \left( i\left[\frac{2\kappa}{\Delta\Gamma} \mp \sqrt{\frac{4\kappa^2}{(\Delta\Gamma)^2} - 1}\right], \ 1 \right)^T. \tag{3}$$

When $\Delta\Gamma$ equals $2|\kappa|$, the two eigenstates coalesce to give an EP with a square-root singularity. Here, $\Delta\Gamma$ and $\kappa$ are the two key external parameters.

Such an EP can be achieved by keeping the coupling constant fixed at $\kappa = \kappa_0$ while varying the additional loss monotonically, with $\Delta\Gamma/2 = \kappa_0(1+\xi)$, where $\xi$ is a tunable parameter. In terms of $\xi$, the two eigenfrequencies have the forms $\tilde{\omega}_\pm = \omega_0 - i\Gamma_{av} \pm \sqrt{2}|\kappa_0|\sqrt{-\xi}$. Varying $\xi$ from a negative to a positive value drives the system from an exact phase through an EP at $\xi = 0$ to a broken phase. Similarly, the same EP can be reached by keeping the additional loss at a constant value $\Delta\Gamma = \Delta\Gamma_0$ while varying the coupling constant $\kappa$. If $\kappa$ is a linear function of another parameter $\eta$ near the EP, with $\kappa = \frac{\Delta\Gamma_0}{2}(1-\eta)$, the eigenfrequencies become $\tilde{\omega}_\pm = \omega_0 - i\Gamma_{av} \pm \left|\frac{\Delta\Gamma_0}{\sqrt{2}}\right|\sqrt{-\eta}$. The same square-root singular behavior near the EP is obtained when the system is driven from an exact phase with $\eta < 0$ to a broken phase with $\eta > 0$. In this simple case, the EP is isotropic in the 2D parameter space of $(\xi, \eta)$. This type of square-root EP is commonly found in various non-Hermitian systems.

We now consider the case when the coupling constant changes quadratically instead of linearly near the EP, as shown schematically in Fig. 1(a) and described by the following equations:

$$\kappa = \kappa_0(1-\eta^2), \tag{4}$$

$$\Delta\Gamma/2 = \kappa_0(1+\xi). \tag{5}$$

Equations (4) and (5) give the following eigenfrequencies and eigenvectors as functions of $\xi$ and $\eta$:

$$\tilde{\omega}_\pm = \omega_0 - i\Gamma_{av} \pm i\sqrt{2}|\kappa_0|\sqrt{\eta^2+\xi}, \tag{6}$$

$$\tilde{\phi}_\pm^R = \left(i\left[1-(\eta^2+\xi)\mp i\sqrt{2(\eta^2+\xi)}\right],\ 1\right)^{\mathrm{T}}. \tag{7}$$

The real and imaginary parts of the eigenfrequencies in the parameter space are shown in Figs. 1(b) and 1(c). We can see from Eq. (6) that when $\xi<0$, the system is in an exact phase when $\eta^2<|\xi|$ and in a broken phase when $\eta^2>|\xi|$. The system possesses two EPs at $\eta=\pm\sqrt{-\xi}$ which have the *same* chirality because they carry the same defective eigenstate $(i,1)^{\mathrm{T}}$ as can be seen from Eq. (7). These two EPs with the same chirality coalescence at $\xi=0$, producing a linear crossing of $\eta$ in the imaginary parts of the eigenfrequencies, as can be seen from Eq. (6) and Figs. 1(c) and 1(e). However, if we fix $\eta$ at zero and vary $\xi$ instead, Eq. (6) shows a typical square-root EP at $\xi=0$. The behavior of the eigenvalues near the EP is illustrated in Figs. 1(d) and 1(e).

In order to confirm the anisotropic behavior, we show in Fig. 1(f) the phase rigidity [29,31], defined as $r_j = \langle \tilde{\phi}_j^R | \tilde{\phi}_j^R \rangle^{-1}$ for each state $j$, near the hybrid EP where $|\tilde{\phi}_j^R\rangle$ are the normalized biorthogonal right eigenvectors. The phase rigidity approaches zero at the hybrid EP with a critical exponent, 1/2 (solid curve), when $\eta$ is fixed at zero and $\xi$ is varied, revealing a typical square-root behavior for a square-root EP. However, when $\xi$ is fixed at zero and $\eta$ is varied, the exponent becomes unity (dashed line). EPs exhibiting such anisotropic behaviors have been called hybrid EPs [35]. The coalescence of two EPs with the same chirality was actually discovered earlier [31] and was called an order-1 EP because the critical exponent of the phase rigidity is unity.

*Experimental realization of a hybrid EP.* We next demonstrate a hybrid EP with acoustic experiments by using coupled acoustic cavities which can realize the exact same non-Hermitian Hamiltonians as those shown in Eq. (1). A photograph of our system is shown in Fig. 2(a). Two identical stainless steel cylindrical cavities are filled with air. Their height and diameter are 120mm and 25mm, respectively. To establish the coupling, the two cavities are connected through an array of 21 equally spaced holes each with a diameter of 3.9mm. The spacing between adjacent holes is 5mm. These holes can be closed by blocking them with Blu-Tack putty, as shown in Fig. 2(a).

We will independently tune both the coupling constant and the additional loss and we need to find a simple way to realize $\eta$ so as to achieve the quadratic coupling behavior $\kappa = \kappa_0 (1-\eta^2)$ required for a hybrid EP. To do this, we use the second eigenmode ($\omega_0 = 2850 \text{Hz}$) of an isolated cavity as the on-site state. Figure 2(a) shows that the pressure profile of this mode is symmetric about the central cross-sectional plane of the cavity with the maximum pressure amplitude found at the center and the two ends of the cavity. Since the coupling constant $\kappa$ is proportional to the pressure amplitude, $\kappa$ can be tuned by choosing which coupling holes to close. And since the eigenmode is symmetric about the central plane of the cavity, the coupling is naturally an even function of $\Delta z$, with the mid-point of the cavity's axis defined as $z=0$.

Experimentally, as shown in Fig. 2(a), we label the holes in the positive $z$ direction from +1 to +10, and the holes in the negative $z$ direction from -1 to -10, whereas the hole at the center is labeled 0. We leave two adjacent holes open at any one time and close the rest to achieve sufficient coupling strength. The pumping loudspeaker is placed inside Cavity A. The resulting pressure spectra measured in Cavity A at different coupling constants are shown in Fig. 2(b). We can see two resonant peaks in each spectrum, due to the anti-crossing of the two on-site modes. By opening holes at positions away from $z=0$, the coupling is reduced as indicated by the narrower splitting between the two peaks. To obtain the value of $\kappa$ for each configuration of coupling holes, we employ the Green's function method to fit the experimentally measured pressure spectrum. The results are shown in Fig. 2(c). In Fig. 2(b), we also show the measured spectra for three different configurations. Using the eigenfrequencies and right/left eigenvectors of the Hamiltonian (1), the Green's function of our system can be defined as [31]

$$\vec{\vec{G}}(\omega) = \sum_{j=1}^{2} \frac{|\tilde{\phi}_j^R\rangle\langle\tilde{\phi}_j^L|}{\omega - \tilde{\omega}_j}, \tag{8}$$

where $|\tilde{\phi}_j^R\rangle$ and $\langle\tilde{\phi}_j^L|$ are the normalized biorthogonal right and left eigenvectors, and $\tilde{\omega}_j$ are the eigenfrequencies. The response function we are interested in is $|P(\omega)| = A|\langle p|\vec{\vec{G}}(\omega)|s\rangle|$, where $|s\rangle$ and $|p\rangle$ are two column vectors describing the source and probe information. For example, in this two-cavity case, the two basis vectors are $(1,0)^T$ for Cavity A and $(0,1)^T$ for Cavity B. According to the experimental setup in Fig. 2(b), both $|s\rangle$ and $|p\rangle$ are $(1,0)^T$. The value of $\kappa_0$ is experimentally determined to be $\kappa_0 = -22.7$Hz when only coupling hole-0 is left open while the rest are all closed, as shown by the blue star in Fig. 2(c).

To see the quadratic relations between $\kappa$ and the locations of holes, we define an effective hole location $z_{eff}$. Owing to the mode profile of the on-site mode which is a cosine function of $z$, we define $z_{eff}$ as $\cos^2\left(\frac{2\pi}{h}z_{eff}\right) = \frac{1}{2}\sum_{i=1}^{2}\cos^2\left(\frac{2\pi}{h}z_i\right)$. Here, $h$ is the height of the cavity and $z_i$ are the actual locations of the coupling holes. We plot the coupling constant $\kappa$ as a function of $z_{eff}$ with red circles in Fig. 2(c). For comparison, we also plot the coupling relation $\kappa = \kappa_0\left[1 - C_k\left(\frac{z_{eff}}{h}\right)^2\right]$ with $\kappa_0 = -22.7$Hz and $C_k = 30.0$ using a solid gray line in Fig. 2(c). Clearly, in our system, $\eta$ corresponds to $\sqrt{C_k}z_{eff}/h$. We can see that when $z_{eff}/h < 0.1$, experimental results agree well with the analytical data. The two deviate only when $z_{eff}/h > 0.1$ as higher order terms become significant when $z_{eff}$ is sufficiently large.

We realize the additional loss $\Delta\Gamma/2 = \kappa_0(1+\xi)$ by monotonically increasing the volume of the dissipative acoustic sponges placed symmetrically at the top and bottom ends of Cavity B. Clearly, $\xi$ increases with the sponge volume. To determine the value of $\Delta\Gamma$, first we need to find $\Gamma_0$ which is determined to be $\Gamma_0$ = 9.35Hz from the measured full width at half maximum (FWHM) of the peak of an isolated cavity without any sponge added, as shown by the gray squares in Fig. 2(d) (set-0). $\Delta\Gamma$ is then obtained by subtracting $\Gamma_0$ from the measured FWHM of the peak with the sponges added (set-1 to set-5; see Fig. 2(d)). The red, blue, magenta, green, and orange circles represent the measured pressure spectra showing increasing values of FWHM with a gradually increasing sponge volume. The results of $\Delta\Gamma$ are 18.16Hz, 30.01Hz, 44.01Hz, 47.32Hz, and 50.91Hz for loss sets 1-5, respectively.

We have now established a way to produce the quadratic coupling behavior $\kappa = \kappa_0(1-\eta^2)$ by choosing the second cavity mode and regarding the effective locations of connecting holes as $\eta$. We have also shown that the additional loss $\Delta\Gamma$ is linearly dependent on the sponge volume, which is represented by the variable $\xi$. We are thus ready to investigate the anisotropic behavior of the hybrid EP by systematically changing the sponge volume and leaving certain coupling holes open while closing the rest. Naturally, the 2D parameter space $(\xi,\eta)$ becomes the $(\Delta\Gamma,z_{eff})$ space in our experiment. We pump Cavity A with a loudspeaker. The sponges are inserted in Cavity B. First, we fix the sponge volume at set-1, and reduce the coupling strength gradually. The coupling is varied in the same way as in the experiment whose results are shown in Fig. 2(b). The pressure spectra measured in Cavity A are shown in Fig. 3(c). We can see two peaks in the spectra and the system is in the exact phase. The reduction in $\kappa$ causes the splitting of these peaks to decrease, pushing the system toward the EP, which is a typical square-root EP phenomenon. Next, we repeat the experiment with an increasing sponge volume in Cavity B. The increases are the same as those shown in Fig. 2(d). The results are shown by markers in Figs. 3(d) and 3(e). Only one peak is seen in Fig. 3(e), as the larger dissipation moves the system toward the broken phase. Note that in the broken phase only the solutions with smaller values of $\text{Im}[\tilde{\omega}]$ can be measured.

In Figs. 3(c-e), the solid curves show the response functions calculated with the Green's function method using the experimentally determined values of $\Delta\Gamma$ and $\kappa$ shown in Fig. 2. The consistency between theoretical and experimental results indicates that the Green's function method captures the physics faithfully. We also use the experimentally determined values of $\Delta\Gamma$ and $\kappa$ to obtain the imaginary parts of the eigenfrequencies using Eq. (2), which are plotted in Fig. 3(a) with solid lines and labeled as "model". From the experimental data in Figs. 3(c-e), we can also extract the imaginary parts of the eigenfrequencies $\text{Im}[\tilde{\omega}]$ from the $(\Delta\Gamma, z_{eff})$ space directly by the Green's function method, and plot the results in Fig. 3(a) with open circles. Again we find good agreement between the analytic formula and the experimental data.

The existence of a hybrid EP and its anisotropic behavior can be seen from Fig. 3(a) as follows. First, we fix $z_{eff} = 0$ and observe that $\text{Im}[\tilde{\omega}]$ bifurcates when $\Delta\Gamma$ increases (from set-1 to set-5), indicating that the system passes through a typical square-root EP, consistent with the solid line shown in Fig. 1(e). However, if we vary the holes to close (i.e. $z_{eff}$) for any fixed value of additional loss $\Delta\Gamma$, then the behavior changes substantially for different loss sets as discussed previously. We see that the $\text{Im}[\tilde{\omega}]$ for loss set-1 and loss set-2 also bifurcates when $|z_{eff}|$ increases. Figure 3(a) shows clearly the existence of a pair of square-root EPs, one for $+z_{eff}$ and the other for $-z_{eff}$. As mentioned before, these two EPs have the same chirality. So when the additional loss increases, these two EPs should coalesce and produce a hybrid EP. For loss set-3, when the coupling strength ($z_{eff}$) is varied, the two square-root EPs in the $\text{Im}[\tilde{\omega}]$ almost coalesce at the point $z_{eff}=0$, as indicated by the gray lines in Fig. 3(b). Furthermore, the dispersion of the $\text{Im}[\tilde{\omega}]$ is almost linear at small $z_{eff}$, indicating that loss set-3 is in close proximity to the hybrid EP. When the additional loss is further increased, the $\text{Im}[\tilde{\omega}]$ exhibits an avoided crossing, and produces a "gap" in the spectrum as shown in Fig. 3(b) for set-4 and set-5. By tracing a line from the two square-root EPs with the same chirality (set-1 and set-2) to the avoided crossings of the eigenfrequencies (set-4 and set-5), we can see that the two bifurcations of the $\text{Im}[\tilde{\omega}]$ move from the $z_{eff}$ axis in the exact phase to the $\Delta\Gamma$ axis in the broken phase, indicating the existence of a transition point — the hybrid EP.

*Conclusion.* We have demonstrated experimentally the existence of a hybrid EP in a system consisting of two coupled cavities. The critical behaviors of the hybrid EP are distinct in two different directions in the parameter space. Along one particular direction, the hybrid EP is the coalescence of two square-root EPs with the same chirality, whereas along the other direction the EP gives rise to two linear dispersions in the imaginary part of the eigenfrequencies. This is different from the linear dispersions found in the real eigenvalues of a diabolic point, such as the Dirac point in graphene, which can be considered as the coalescence of two square-root singularities having opposite chiralities when non-Hermiticity is added to the system.

*Acknowledgements.* This work is supported by the Research Grants Council of Hong Kong (grant no. AoE/P-02/12).

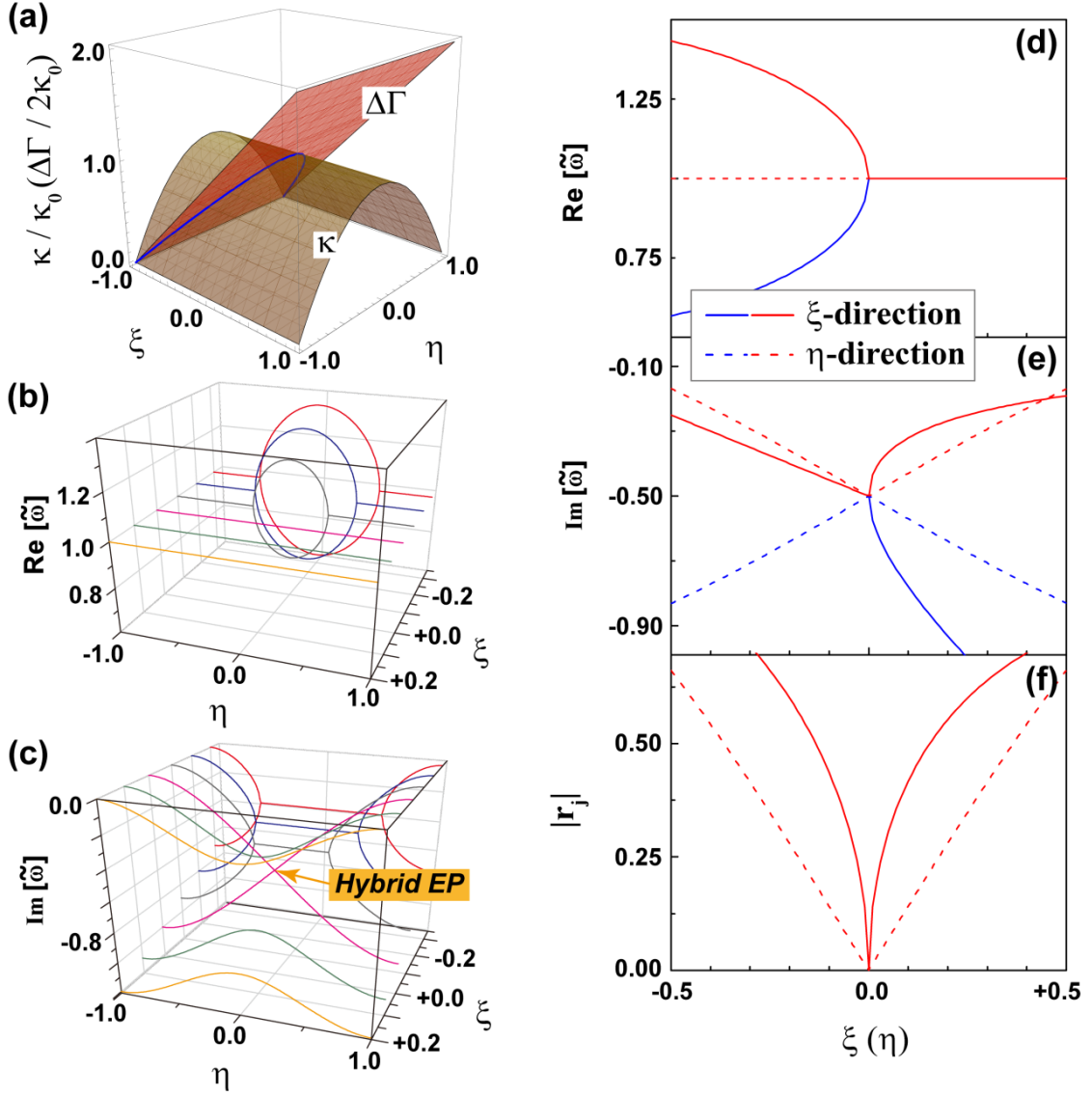

FIG. 1. (color online) (a) The normalized coupling constant $\kappa/\kappa_0$ and normalized loss difference $\Delta\Gamma/(2\kappa_0)$ in the $(\xi,\eta)$ space are shown by the red and brown surfaces, respectively. The solid blue line corresponds to EPs. Real and imaginary parts of the calculated eigenfrequencies in the $(\xi,\eta)$ space are shown in (b) and (c), respectively. (d) Real and (e) imaginary parts of the eigenfrequencies along the $\xi$-direction ($\eta=0$) and $\eta$-direction ($\xi=0$) are shown by solid and dashed lines, respectively. Phase rigidity of the corresponding state along the $\xi$-direction ($\eta=0$) and $\eta$-direction ($\xi=0$) is shown in (f). The parameters used are $\omega_0=1.0$, $\Gamma_0=0.0$,

and $\kappa_0 = 0.5$.

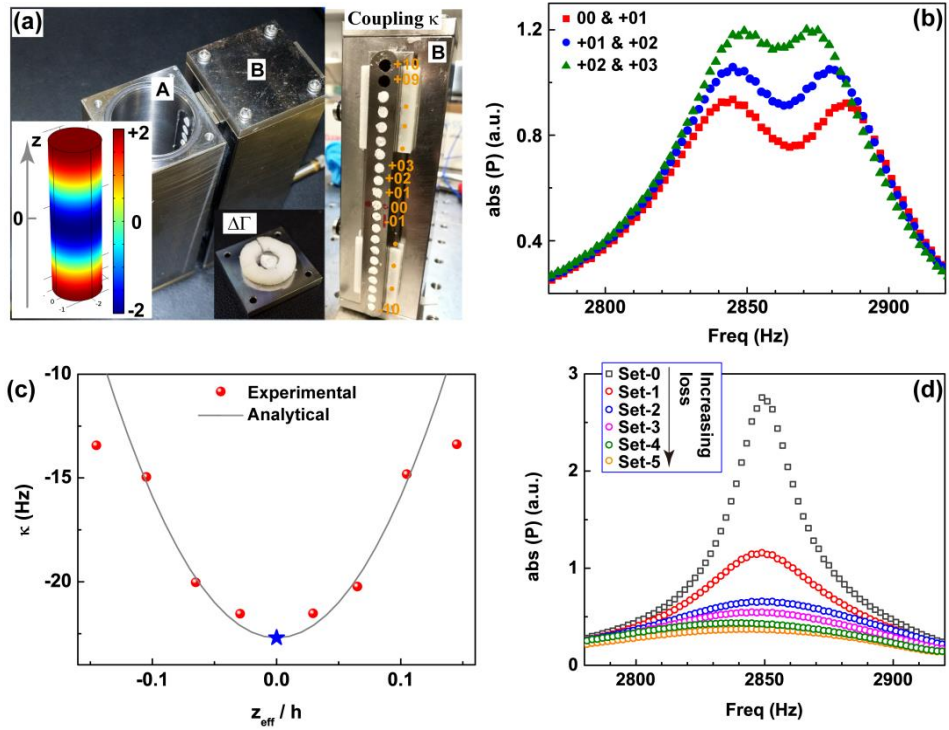

FIG. 2. (color online) (a) Experimental setup. The left panel shows the coupled cavities, and the inset is the simulated pressure distribution of the second resonance mode in the z-direction. The right panel shows the coupling holes in the experiment (in this case the top two holes are left open). The right inset shows one of the sponges used to increase loss. (b) Measured pressure spectra for the two-cavity system with different sets of coupling holes. (c) Coupling strengths as a function of effective hole locations $z_{eff}$ (see text) obtained from experimental data are shown with red dots. The solid gray line shows the analytical fitting to the experimental data. The blue star is obtained from experimental data with only hole-0 open. (d) Measured pressure spectra for a single cavity with tunable loss. The grey squares represent the spectrum without additional loss. The red, blue, magenta, green, and orange circles denote responses to increasing loss. Uncertainties in the measured data are no larger than the size of the markers.

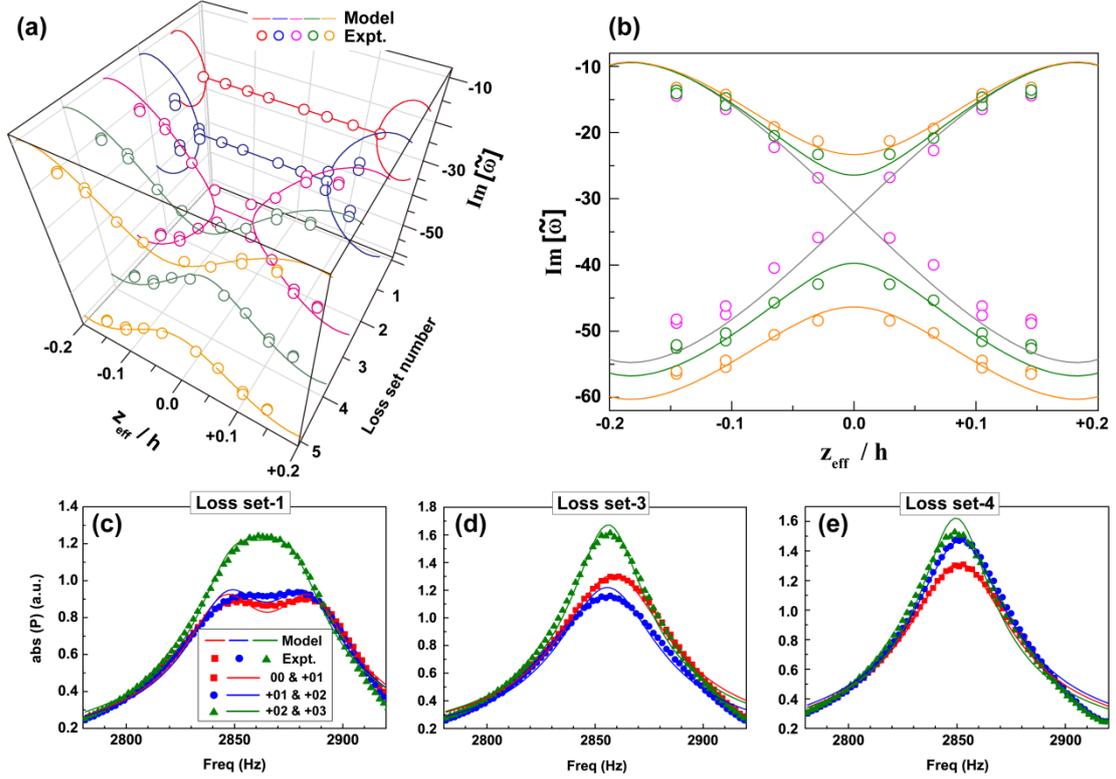

FIG. 3. (color online) (a) Imaginary parts of the eigenfrequencies as functions of $z_{eff}$ and additional loss (set number). Solid lines are calculated using the experimental couplings and losses extracted from Fig. 2, and the open circles are obtained directly from experimental data. (b) Imaginary parts of the eigenfrequencies as a function of $z_{eff}$ of loss set-3, set-4, and set-5 are plotted with magenta, green, and orange circles, respectively. Uncertainties in the measured data are no larger than the size of the markers. The two gray lines show the linear crossing behavior. Measured pressure spectra for three different loss sets are plotted in (c), (d), and (e). All solid lines are calculated using the Green's function method and the experimental couplings and losses extracted from Fig. 2.

16